\documentclass{aa}
\usepackage{amssymb}
\usepackage{amsmath}
\usepackage{graphicx}
\usepackage{txfonts}

\newcommand{\calE}{{\mathcal{E}}}
\newcommand{\calN}{{\mathcal{N}}}
\newcommand{\bfk}{{\boldsymbol{k}}}
\newcommand{\bfr}{{\boldsymbol{r}}}
\newcommand{\bfv}{{\boldsymbol{v}}}
\newcommand{\txd}{{\text{d}}}
\newcommand{\txp}{{\text{p}}}

\begin{document}

\title{The dynamical structure of isotropic spherical galaxies with
a central black hole}

\author{%
Maarten Baes\inst{1,2} 
\and 
Herwig Dejonghe\inst{1} 
\and 
Pieter Buyle\inst{1}
}

\titlerunning{Isotropic spherical galaxies with a central black hole}

\authorrunning{M. Baes et al.}

\institute{
Sterrenkundig Observatorium, Universiteit Gent, Krijgslaan
281-S9, B-9000 Gent, Belgium
\and
European Southern Observatory, Casilla 19001, Santiago, Chile
}

\offprints{Maarten Baes, \email{mbaes@eso.org}}

\date{Printed \today}

\abstract{We discuss the kinematical structure of a two-parameter
family of isotropic models with a central black hole. The family
contains the slope of the central density cusp and the relative black
hole mass as parameters. Most of the basic kinematical quantities of
these models can be expressed analytically. This family contains three
distinct models where also the distribution function, differential
energy distribution and spatial LOSVDs can be expressed completely
analytically. Each of these models show a drastically different
behaviour of the distribution function. Although the effect of a black
hole on the distribution function is very strong, in particular for
models with a shallow density cusp, the differential energy
distribution is only marginally affected. We discuss in detail the
effects of a central black hole on the LOSVDs. The projected velocity
dispersion increases with black hole mass at small projected radii,
but the effect of a black hole on the shape of the LOSVDs
(characterized by the $h_4$ parameter) is less straightforward to
interpret. Too much reliance on the wings of the LOSVDs and the value
of the $h_4$ parameter to determine black hole masses might hence be
dangerous.  \keywords{ black hole physics -- galaxies: kinematics and
dynamics -- galaxies: structure } }

\maketitle

\section{Introduction}

During the past decade, various observational discoveries have changed
our view on galactic nuclei rather drastically. Firstly, galactic
nuclei are generally observed to have a cuspy density distribution at
small radii, with densities behaving as $\rho(r)\propto
r^{-\gamma}$. The observed cusp slopes are far from uniform and range
from zero to more than 2 (Lauer et al.~1995; Gebhardt et al.~1996;
Ravindranath et al.~2001; Seigar et al.~2002; Genzel et al.~2003;
Scarlata et al.~2004). Secondly, high-resolution imaging has revealed
that a substantial fraction of the galactic nuclei in both spiral and
elliptical galaxies show small-scale structure, in the form of bars,
mini-spirals or dust lanes (Phillips et al.~1996; Malkan, Gorjian \&
Tam~1998; Carollo, Stiavelli \& Mack~1998; Tran et al.~2001; Martini
et al.~2003).  Thirdly, there is now enough credible evidence that
(nearly) all nearby galaxies harbor a supermassive black hole in their
centre. Intriguingly, the masses of these putative black holes are
tightly coupled to large-scale parameters of the host galaxies
(Kormendy \& Richstone~1995; Ferrarese \& Merritt~2000; Gebhardt et
al.~2000; Graham et al.~2001; McLure \& Dunlop~2002; Ferrarese~2002;
Baes et al.~2003a; Marconi \& Hunt~2003). Clearly, these various
aspects are not isolated features, they all play part in the processes
which shape galactic nuclei and galaxies in general. Unfortunately, we
still know little about the formation processes and evolutionary
scenarios of galactic nuclei and their black holes. For example, it is
still unclear when and how supermassive black hole are being fuelled
and what the mutual influence is between black holes and the central
density cusps in galactic nuclei. Various relevant processes probably
play at scales which are beyond (or at the limit of) the current
observational capabilities. In order to increase our understanding,
more detailed theoretical modelling, for example using N-body of SPH
simulations, is necessary.

As a starting point or general framework for such studies, one needs a
set of reference models which are simple enough and still can present
a wide enough variety in structural characteristics. Because of the
observed cuspy nature of galaxy centers, well explored models such as
the Plummer model (Plummer~1911; Dejonghe~1987) and the isochrone
sphere (H\'enon 1959, 1960) are less suitable. Scale free models are a
useful alternative. Such models have a density profile that decreases
as $r^{-\gamma}$, and they include the singular isothermal sphere as a
special case. Due to the simplicity of these models, their dynamical
structure can be easily studied, even in axisymmetric or triaxial
generalizations (e.g.\ Toomre~1982; Evans~1994; Qian et al.~1995; de
Bruijne et al.~1996; Evans et al.~1997; Jalali \& de
Zeeuw~2001). Unfortunately, scale free models always have an infinite
total mass, with the mass diverging in the centre for $\gamma\geq3$
and at large radii for $\gamma\leq3$.

A set of models that does not suffer from this disadvantage is the
family of $\gamma$-models, introduced independently by Dehnen~(1993)
and Tremaine et al.~(1994). These models have a central $r^{-\gamma}$
density cusp ($\gamma<3$) and the density falls at $r^{-4}$ at large
radii, such that the total mass is always finite. Special cases are
the well-known Hernquist and Jaffe models (Hernquist 1990; Jaffe
1984).  Many interesting dynamical properties such as the intrinsic
and projected velocity dispersions, the distribution function and the
differential energy distribution can be calculated analytically for
the $\gamma$-models, under the assumption of a self-consistent
isotropic dynamical structure. Extensions towards an anisotropic
distribution functions or flattened geometry have been presented
(e.g.\ Dehnen \& Gerhard~1994; Hiotelis 1994; Carollo et al.~1995;
Zhenglu~2000; Baes et al.~2002a; Zhenglu \& Moss~2002).

As supermassive black holes appear to the present in nearly all
galactic nuclei, it would be very interesting to extend the family of
$\gamma$-models with a central black hole. In principle this is quite
straightforward: one just has to add an extra contribution from the
black hole to the stellar potential and re-calculate the dynamical
properties with this new potential. In fact, Tremaine et al.~(1994)
consider the case where a central black hole is present in the
$\gamma$-models. They present analytical expressions for the velocity
dispersions and discuss the effect of a black hole on the distribution
function. Unfortunately, they do not seek for analytical expressions
for the distribution function and they leave some of the most
interesting kinematical properties such as the differential energy
distribution and the LOSVDs undiscussed. Ciotti~(1996) provides a
first attempt to construct completely analytical dynamical models for
galaxies with a central black hole. Extending the work by Carollo et
al.~(1995), he considers a set of Hernquist models embedded in a dark
matter halo. By setting the dark halo radius to zero this halo reduces
to a central black hole. He demonstrates that many of the interesting
dynamical quantities, including the distribution function and the
differential energy distribution, can be calculated analytically
(albeit as rather complicated functions involving Jacobian
functions). Baes \& Dejonghe~(2004) present a one-parameter family
with a steep $\gamma=\tfrac{5}{2}$ cusp slope and the black hole mass
as a parameter. Almost all interesting properties of this family of
models can be written in terms of elementary functions.

In this paper, we present a detailed analysis of a two-parameter
family of spherical isotropic dynamical models based on the
$\gamma$-models. The family contains as parameters the slope of the
central density cusp $\gamma$ and the ratio $\mu$ of the central black
hole mass to the total mass of the system. The parameter space covered
by these models goes from weakly cusped models to very centrally
concentrated models with an infinite stellar potential well, and from
self-consistent models without black hole to systems where the
dynamical structure is completely dominated by the central black
hole. In Section~2 we define the models. In Section~3 we derive some
basic properties, most of which can be calculated completely
analytically. In Section~4 we discuss the energy budget of the models
and look at the virial theorem. In Section~5 we derive expressions for
and discuss the distribution function and the differential energy
distribution. In Section~6 we study the LOSVDs of the models in our
family and discuss the observational signature of a black hole. In the
last Section we summarize the results, and in the Appendices we
present some mathematical expressions which might be useful for people
who wish to use these models as input for further theoretical studies.

\section{Definition of the models}

The $\gamma$-models have a luminosity density
\begin{equation}
    \rho(r)
    =
    \frac{3-\gamma}{4\pi}\,
    \frac{1}{r^\gamma(1+r)^{4-\gamma}},
\label{rho}
\end{equation}
The parameter $\gamma$ determines the density slope of the system at
small radii; it can assume values between 0 and 3. All models have a
luminosity density that behaves as $r^{-4}$ at large radii, such that
the total luminosity is finite.

The gravitational potential of the models we consider is the sum of
two contributions: the stellar mass and a central black hole. We
introduce the parameter $\mu$ as the relative importance of the black
hole mass to the {\em total} mass in the system, such that $\mu$ can
assume values only between 0 and 1. Note that the convention we use is
similar to the one adopted by Baes \& Dejonghe~(2004), but is
different from the convention used in e.g.~Tremaine et al.~(1994) and
Zhao~(1996). In these papers, $\mu$ denotes the black hole mass
relative to the {\em stellar} mass and the normalization is such that
the stellar mass is set to unity. We prefer to set the {\em total}
mass of the galaxy equal to unity however, because all models then
have the same behaviour at large radii.

In the limit $\mu=0$, there is no black hole and we recover the
self-consistent $\gamma$-models described in detail by Dehnen~(1993)
and Tremaine et al.~(1994). The potential reduces to the stellar
potential
\begin{equation}
    \Psi(r)
    \rightarrow
    \Psi_*(r)
    =
    \frac{1}{2-\gamma}
    \left[1-\left(\frac{r}{1+r}\right)^{2-\gamma}\right].
\label{potnoBH}
\end{equation}
This expression is not valid for $\gamma=2$, the Jaffe model. If we
take the limit $\gamma\rightarrow2$ in the previous expression we
obtain the Jaffe potential $\Psi_*(r) = \ln(1+1/r)$. This special case
divides the family of $\gamma$-models in two classes: self-consistent
$\gamma$-models with $\gamma<2$ have a finite potential well with
depth $\Psi_0=1/(2-\gamma)$, whereas the stellar potential well is
infinitely deep for the models with $2\leq\gamma<3$.

The other extreme case on the range of possible black hole masses is
$\mu=1$, corresponding to systems where the entire mass resides within
the central black hole. In this case, the total gravitational
potential reduces to a Kepler potential,
\begin{equation}
    \Psi(r)
    \rightarrow
    \Psi_\bullet(r)
    =
    \frac{1}{r}.
\label{potdomBH}
\end{equation}
Although these systems where the dynamics are dominated by a central
black hole do not form a realistic representation of real galaxies,
they are useful in order to investigate the maximal effect of a black
hole on the dynamical properties of the $\gamma$-models.

The general case $0<\mu<1$ is intermediate between these two extreme
cases. We can write the cumulative mass function and the potential of
the $\gamma$-models as
\begin{gather}
    M(r)
    =
    (1-\mu)
    \left(\frac{r}{1+r}\right)^{3-\gamma}
    +
    \mu,
    \label{M}
    \\
    \Psi(r)
    =
    \frac{1-\mu}{2-\gamma}
    \left[1-\left(\frac{r}{1+r}\right)^{2-\gamma}\right]
    +
    \frac{\mu}{r}.
    \label{pot}
\end{gather}
The two equations (\ref{rho}) and (\ref{pot}) form a two-parameter
family of potential-density pairs. Each pair of parameters
$(\gamma,\mu)$ completely defines a dynamical model. The study of
these models as a function of the parameters $\gamma$ and $\mu$ is the
goal of this paper.

\section{Basic properties}

\subsection{The surface brightness}

The surface brightness profile of a $\gamma$-model can be found by
projecting the luminosity density on the plane of the sky,
\begin{equation}
    I(R)
    =
    2\int_R^\infty
    \frac{\rho(r)\,r\,\txd r}{\sqrt{r^2-R^2}}.
\end{equation}
At large radii, the surface brightness falls as $R^{-3}$. In the
central regions, models with $\gamma<1$ have a finite central surface
brightness (in spite of the divergence of the luminosity density),
whereas models with $\gamma>1$ have a surface brightness profile that
diverges as $R^{1-\gamma}$. The surface brightness can be evaluated
analytically for all integer and half-integer values of
$\gamma$. Analytical expressions and other photometric quantities such
as the cumulative surface brightness and the half-light radius can be
found in Hernquist~(1990), Dehnen~(1993), Tremaine et al.~(1994) and
Baes \& Dejonghe~(2004).

\subsection{The velocity dispersion}

For a spherical isotropic system, the intrinsic velocity dispersion
profile can be found using the solution of the lowest-order Jeans
equation (Dejonghe~1986; Binney \& Tremaine 1987),
\begin{equation}
    \sigma^2(r)
    =
    \frac{1}{\rho(r)}\int_r^\infty \frac{M(r)\,\rho(r)\,\txd r}{r^2}.
\label{jeans}
\end{equation}
After substitution of the general cumulative mass function~(\ref{M})
and some algebra, we obtain the expression
\begin{equation}
    \rho(r)\sigma^2(r)
    =
    (1-\mu)W_\gamma(r) + 2\mu\,W_{\frac{3+\gamma}{2}}(r),
\label{sig}
\end{equation}
where the function $W_\gamma(r)$ is defined as
\begin{equation}
    W_\gamma(r)
    =
    \frac{3-\gamma}{4\pi}
    \int_r^\infty
    \frac{r^{1-2\gamma}\txd r}{(1+r)^{7-2\gamma}}.
\label{defW}
\end{equation}
From equation~(\ref{sig}) we see that $W_\gamma(r)$ is nothing but the
velocity dispersion $\rho(r)\sigma_*^2(r)$ of the self-consistent
$\gamma$-model, which can be evaluated in terms of elementary
functions for all values of $\gamma$ (see Appendix~A).

\subsection{The projected velocity dispersion}

The projected velocity dispersion profile $\sigma_\txp(R)$ can be
found by projecting the intrinsic dispersion on the plane of the sky,
\begin{equation}
    \sigma_\txp^2(R)
    =
    \frac{2}{I(R)}
    \int_R^\infty
    \frac{\rho(r)\sigma^2(r)\,r\,\txd r}{\sqrt{r^2-R^2}}.
\label{nupsigmap2}
\end{equation}
This expression can be written in a form very similar to
equation~(\ref{sig}),
\begin{equation}
    I(R)\sigma_\txp^2(R)
    =
    (1-\mu)Y_\gamma(R) + 2\mu\,Y_{\frac{3+\gamma}{2}}(R),
\label{sigbullet}
\end{equation}
with (Tremaine et al.~1994)
\begin{equation}
    Y_\gamma(R)
    =
    \frac{3-\gamma}{2\pi}
    \int_R^\infty
    \frac{r^{1-2\gamma}\,\sqrt{r^2-R^2}\,\txd r}{(1+r)^{7-2\gamma}}.
\label{defY}
\end{equation}
The function $Y_\gamma(R)$ represents the projected velocity
dispersion $I(R)\sigma^2_{\txp,*}(r)$ of the self-consistent
$\gamma$-model. It can be expressed analytically for all integer and
half-integer values of $\gamma$, and has to be calculated numerically
for the other values of $\gamma$. A useful transformation for the
numerical calculation of the integral can be found in Appendix~B of
Dehnen~(1993).

\section{The energy budget and the virial theorem}

The (scalar) virial theorem states that any steady-state system
satisfies the relation
\begin{equation}
  2K
  =
  U,
\end{equation}
where $K$ is the total kinetic energy and $U=-W$ is the total
(binding) potential energy of the system. It is easy to verify that
the $\gamma$-models without a central black hole satisfy the virial
theorem. For a self-consistent spherical system the total potential
and kinetic energy can be found as
\begin{align}
    U_*
    &=
    2\pi\int_0^\infty
    \rho(r)\,\Psi_*(r)\,r^2\,\txd r.
    \label{U*}
    \\
    K_*
    &=
    6\pi\int_0^\infty
    \rho(r)\sigma_*^2(r)\,r^2\,\txd r
    \label{K1}
    \\
    &=
    2\pi\int_0^\infty \rho(r)\,M_*(r)\,r\,\txd r,
    \label{K2}
\end{align}
where the second expression for the kinetic energy is derived by
substituting equation (\ref{jeans}) into the first expression and
partial integration. Through substitution of the
expressions~(\ref{rho}), (\ref{potnoBH}) and (\ref{M}) into these
formulae, one finds that the self-consistent $\gamma$-models have an
infinite energy budget for $\gamma\geq\tfrac{5}{2}$, whereas for
$\gamma<\tfrac{5}{2}$ we obtain (Tremaine et al.~1994)
\begin{equation}
    U_*
    =
    2K_*
    =
    \frac{1}{5-2\gamma}.
\end{equation}
For systems with a central black hole, the total kinetic energy can
still be calculated using equation (\ref{K1}) or (\ref{K2}), with the
contribution of the black hole taken into account in the dispersion or
the cumulative mass distribution. The total potential energy, however,
{\em cannot} be found by just replacing the stellar potential by the
total potential, as the expression~(\ref{U*}) is derived under the
assumption that the system is self-consistent. The correct expression
for the potential energy consists of an internal and an external
contribution (Binney \& Tremaine 1987, problem 8.2),
\begin{equation}
  U 
  = 
  \frac{1}{2}
  \iiint \rho(\bfr)\,\Psi_{\text{int}}(\bfr)\, 
  \txd\bfr
  + 
  \iiint \rho(\bfr)\,
  \left|\bfr\cdot\nabla\Psi_{\text{ext}}(\bfr)\right| 
  \txd\bfr.
\end{equation}
In our present case, the external potential is the gravitational potential from the central black hole, such that these formulae reduce to 
\begin{equation}
  U
  =
  2\pi\,(1-\mu)
  \int_0^\infty \rho(r)\,\Psi_*(r)\,r^2\,\txd r
  +
  4\pi\mu \int_0^\infty \rho(r)\,r\,\txd r,
\end{equation}
or equivalently,
\begin{equation}
  U
  =
  2\pi\int_0^\infty \rho(r)\,\Psi(r)\,r^2\,\txd r
  +
  2\pi\mu\int_0^\infty \rho(r)\,r\,\txd r.
\end{equation}
Comparing this expression to the equivalent formula (\ref{U*}) of the
self-consistent models, we notice that the black hole mass adds an
extra contribution to the potential energy apart from its contribution
to the potential of the system. Using the expressions (\ref{rho}),
(\ref{pot}) and (\ref{M}), one finds that the total energy budget is
infinite for $\gamma\geq2$, whereas for $\gamma<2$ we obtain
\begin{equation}
    U
    =
    2K
    =
    \dfrac{1}{5-2\gamma}
    \left[1 + \mu\left(\frac{3-\gamma}{2-\gamma}\right)\right].
\end{equation}
The virial theorem is thus satisfied.

\section{The distribution function and the differential energy distribution}

All the kinematical information on a given system is contained in the
distribution function $f(\bfr,\bfv)$, which represents the number
density of stars in six-dimensional phase space. For spherical
isotropic systems, the distribution function depends only on the
binding energy $\calE = \Psi(r)-v^2/2$. The key to calculating the
distribution function $f(\calE)$ of isotropic spherical models is the
augmented density $\tilde\rho(\Psi)$, i.e.\ the luminosity density
written as a function of the potential. The Eddington formula
specifies how the distribution function can be calculated from the
augmented density,
\begin{align}
        f(\calE)
        &=
        \frac{1}{\sqrt{8}\pi^2}\,
        \frac{\txd}{\txd\calE}
        \int_0^\calE \frac{\txd\tilde{\rho}}{\txd\Psi}\,
        \frac{\txd\Psi}{\sqrt{\calE-\Psi}}
\label{eddington}
\\
        &=
        \frac{1}{\sqrt{8}\pi^2}
        \int_0^\calE \frac{\txd^2\tilde{\rho}}{\txd\Psi^2}\,
        \frac{\txd\Psi}{\sqrt{\calE-\Psi}}
        +
        \frac{1}{\sqrt{8\calE}\,\pi^2}
        \left(\frac{\txd\tilde{\rho}}{\txd\Psi}\right)_{\Psi=0}.
\label{eddington2}
\end{align}
The distribution function obviously is an important characteristic of
the dynamical structure of galaxies.  However, it is not
straightforward to physically interpret the meaning of the
distribution function when expressed as a function of binding
energy. A more natural diagnostic quantity is the differential energy
distribution (DED) ${\mathcal{N}}(\calE)$, which describes the number
of stars per unit binding energy. For isotropic systems, the DED is
simply the product of the distribution function $f(\calE)$ and the
density of states function $g(\calE)$, defined as the phase space
volume accessible for a star with binding energy $\calE$. This
function can be calculated as
\begin{equation}
    g(\calE)
    =
    16\sqrt{2}\,\pi^2
    \int_{\calE}^{\Psi_0}
    \left|r^2\frac{\txd r}{\txd\Psi}\right|\,
    \sqrt{\Psi-\calE}\,\txd\Psi,
\label{dos}
\end{equation}
and depends only on the potential of the system, not on the density
profile (Binney \& Tremaine~1987).

\subsection{Models without a central black hole $(\mu=0)$}

For $\gamma$-models without a black hole, we can calculate the
distribution function $f_*(\calE)$ directly by substituting the
equations~(\ref{rho}) and (\ref{potnoBH}) into the Eddington
relation~(\ref{eddington}). We obtain the expression
\begin{equation}
    f_*(\calE)
    =
    \frac{3-\gamma}{16\sqrt{2}\pi^3}\,
    \frac{\txd}{\txd\calE}
    \int_0^{\calE}
    \frac{(1-t)^3\,[(4-\gamma)\,t+\gamma]\,\txd\Psi_*}
    {t^2\,\sqrt{\calE-\Psi_*}},
\label{fnoBH1}
\end{equation}
where $t\equiv t(\Psi_*)$ is defined as
\begin{equation}
    t(\Psi_*)
    =
    \begin{cases}
    [1-(2-\gamma)\,\Psi_*]^{1/(2-\gamma)}
    &
    \quad\text{if $\gamma\neq2$,}
    \\
    \text{e}^{-\Psi_*}
    &
    \quad\text{if $\gamma=2$.}
    \end{cases}
\label{t}
\end{equation}
This integral can generally be expressed in terms of hypergeometric
functions,
\begin{align}
    f_*(\calE)
    &=
    \frac{3-\gamma}{\sqrt{8}\,\pi^3}\,
    \sqrt{\calE}
    \Bigl[
    (4-\gamma)\,
    {}_2F_1
    \left(
    1,\tfrac{-\gamma}{2-\gamma};\tfrac{3}{2};(2-\gamma)\calE
    \right)
    \nonumber \\
    &\qquad
    -2\,(3-\gamma)\,
    {}_2F_1
    \left(
    1,\tfrac{1-\gamma}{2-\gamma};\tfrac{3}{2};(2-\gamma)\calE
    \right)
    \nonumber \\
    &\qquad
    +2\,(1-\gamma)\,
    {}_2F_1\left(
    1,\tfrac{3-\gamma}{2-\gamma};\tfrac{3}{2};(2-\gamma)\calE
    \right)
    \nonumber \\
    &\qquad
    +\gamma\,
    {}_2F_1\left(
    1,\tfrac{4-\gamma}{2-\gamma};\tfrac{3}{2};(2-\gamma)\calE
    \right)
    \Bigr].
\label{fnoBH}
\end{align}
For $\gamma=2\pm1/n$ with $n$ a natural number, the distribution
function can be written as a combination of elementary functions. In
particular, the self-consistent $\gamma$-models with $\gamma=1$, $\tfrac{3}{2}$
and $\tfrac{5}{2}$ have fairly simple distribution functions (Hernquist~1990;
Dehnen~1993; Tremaine et al.~1994; Baes \& Dejonghe~2004). The
$\gamma=2$ model is a particular case: its distribution function can
be expressed most conveniently in terms of the error function and
Dawson's integral (Jaffe~1983).

The calculation of $g(\calE)$ is also straightforward for the
self-consistent $\gamma$-models.  Substituting the
potential~(\ref{pot}) in equation~(\ref{dos}) one immediately obtains
(Dehnen~1993)
\begin{equation}
    g_*(\calE)
    =
    16\sqrt{2}\pi^2
    \int_{\calE}^{\Psi_0}
    \frac{t^{1+\gamma}\,\sqrt{\Psi_*-\calE}\,\txd\Psi_*}
    {(1-t)^4},
\end{equation}
with $t \equiv t(\Psi_*)$ as in equation~(\ref{t}). The integral in
this equation is quite similar to the integral in
equation~(\ref{fnoBH1}). A closed expression for $g_*(\calE)$ for
general $\gamma$ in terms of hypergeometric functions cannot be
obtained, but the integral can be expressed analytically for all
$\gamma=2\pm1/n$ with $n$ a natural number. Examples of such closed
expressions can be found in Dehnen~(1993) and Baes \& Dejonghe~(2004).

\subsection{Models dominated by a central black hole $(\mu=1)$}

For models where the potential is completely dominated by the central
black hole, the distribution function can also be calculated by a
straightforward application of Eddington's formula. The result can
generally be expressed in terms of hypergeometric functions,
\begin{align}
    f_\bullet(\calE)
    &=
    \frac{4\sqrt{2}(3-\gamma)}{315\pi^3}
    \calE^{5/2}
    \Bigl[
    63\,
    {}_2F_1 \left(3,6-\gamma;\tfrac{7}{2};-\calE\right)
    \nonumber \\
    &\qquad
    -36(1-\gamma)\,\calE\,
    {}_2F_1 \left(3,6-\gamma;\tfrac{9}{2};-\calE\right)
    \nonumber \\
    &\qquad
    -4\gamma(1-\gamma)\,\calE^2
    {}_2F_1 \left(3,6-\gamma;\tfrac{11}{2};-\calE\right)
    \Bigr].
\label{dfdomBH}
\end{align}
For half-integer or integer values of $\gamma$, the distribution
function can be written in terms of elementary functions. In the
central regions of the galaxy ($\calE\rightarrow\infty$), this
distribution function has the asymptotic behaviour
\begin{equation}
    f_\bullet(\calE)
    \sim
    \frac{3-\gamma}{2(2\pi)^{5/2}}\,
    \frac{\Gamma(\gamma+1)}{\Gamma(\gamma-\tfrac{1}{2})}\,
    \calE^{\gamma-3/2}.
\label{poiuy}
\end{equation}
It is interesting to compare these results with those obtained by de
Bruijne et al.~(1996). They consider a set of axisymmetric cuspy
densities in a spherical potential and present two different
analytical families of three-integral distribution functions. For the
special case of isotropic spherical galaxies dominated by a black hole
potential ($q=\delta=1$ and $\beta=0$) the expression~(\ref{poiuy}) is
recovered, after correction for the different normalization
conventions.

Since the density of states function only depends on the potential of
the system, we recover the simple and well-known expression
$g_\bullet(\calE) = \sqrt{2}\pi^3\,\calE^{-5/2}$, independent of
$\gamma$.

A particularly interesting case is the model with $\gamma=\tfrac{1}{2}$. This
model has a simple distribution function and differential energy
distribution,
\begin{gather}
    f_\bullet(\calE)
    =
    \frac{\sqrt{8}}{\pi^3}\,
    \frac{\calE^{5/2}}{(1+\calE)^5},
    \\
    \calN_\bullet(\calE)
    =
    \frac{4}{(1+\calE)^5}.
\end{gather}
For $\gamma$-models with a central black hole, this model is the one
with the smallest possible cusp slope. Indeed, $\gamma$-models with
$\gamma<\tfrac{1}{2}$ cannot support a central black hole when they have an
isotropic dynamical structure, because stars at each binding energy
level will induce a $r^{-1/2}$ cusp at small radii (Tremaine et
al.~1994).

\subsection{The general case $(0<\mu<1)$}

For the general case, a direct calculation of the augmented density is
not the best way to calculate the distribution function. Instead, we
use an approach based on the analysis of Ciotti~(1996), who showed a
convenient way to calculate the distribution function of a set of
two-component models. If we invert the relation (\ref{potnoBH}) to
$r=r(\Psi_*)$, and substitute it into the expression (\ref{pot}), we
obtain
\begin{equation}
        \Psi
        =
        \omega(\Psi_*)
        \equiv
        (1-\mu)\,\Psi_* + \frac{\mu\,(1-t)}{t},
\label{omega}
\end{equation}
with $t\equiv t(\Psi_*)$ as in expression~(\ref{t}). The
relation~(\ref{omega}) links the stellar potential $\Psi_*$ at a given
position in the system to the total potential $\Psi$. We can also
regard it more generally as the definition of a mapping function
$\omega$, which maps the interval $[0,\Psi_0]$ onto the positive real
axis.  It is a monotonically increasing function, and therefore the
inverse function $\omega^{-1}$ exists. Transforming the Eddington
formula (\ref{eddington}) to an integration with respect to the
stellar potential we obtain
\begin{subequations}
\begin{equation}
    f(\calE)
    =
    \frac{1}{\sqrt{8}\pi^2}\,
    \left[\frac{\txd\omega}{\txd\calE_*}(\calE_*)\right]^{-1}
    \frac{\txd{\mathcal{Q}}}{\txd\calE_*}
    (\calE_*)
\label{f1}
\end{equation}
with $\calE_*=\omega^{-1}(\calE)$ and ${\mathcal{Q}}$ a function
defined by the integral
\begin{equation}
    {\mathcal{Q}}(x)
    =
    \int_0^x
    \frac{\txd\tilde\rho}{\txd\Psi_*}\,
    \frac{\txd\Psi_*}{\sqrt{\omega(x)-\omega(\Psi_*)}}.
\label{defQ}
\end{equation}
\end{subequations}
This expression is similar to the formula used by Ciotti~(1996) to
calculate the distribution function of his two-component Hernquist
models.  Whether this integration can be performed analytically
depends on the complexity of both factors of the integrand. It appears
that this integration can be performed analytically (only) for the
models with $\gamma=1$, $\tfrac{3}{2}$ and $\tfrac{5}{2}$. For these
three models, both the augmented density $\tilde\rho(\Psi_*)$ and the
mapping function $\omega(\Psi_*)$ are rational functions. For
$\gamma=1$ and $\gamma=\tfrac{3}{2}$, the factor under the square root
can be reduced to a cubic polynomial, while for $\gamma=\tfrac{5}{2}$
it can be reduced to a quadratic polynomial in $\Psi_*$. As a result,
the function ${\mathcal{Q}}(x)$ and the distribution function can be
expressed by means of elliptic integrals for the former two models,
and completely in terms of elementary functions for the latter
model. More details can be found in Appendix~B.

For other values of $\gamma$ (including $\gamma=2$), an analytical
evaluation of the integral~(\ref{defQ}) is not possible, and the
distribution function has to be calculated numerically. For this goal,
the expression (\ref{f1}) is not particularly useful, because it
involves a differentiation of a numerically determined function. A
more convenient formula for numerical integration can be obtained by
using the alternative form~(\ref{eddington2}) of the Eddington
equation. The second term in this expression vanishes for all
$\gamma$-models, because $\tilde\rho(\Psi) \propto \Psi^4$ at large
radii. If we do the substitution $\Psi\rightarrow\Psi_*$ in this last
equation, we obtain
\begin{equation}
        f(\calE)
        =
        \frac{1}{\sqrt{8}\pi^2}
        \int_0^{\calE_*}
        \frac{\txd}{\txd\Psi_*}
        \left[
        \frac{\txd\tilde\rho}{\txd\Psi_*}
        \Bigl/
        \frac{\txd\omega}{\txd\Psi_*}
        \right]
        \frac{\txd\Psi_*}{\sqrt{\omega(\calE_*)-\omega(\Psi_*)}}.
\label{f2}
\end{equation}
Combined with a numerical solution of the equation
$\calE_*=\omega^{-1}(\calE)$, this formula allows to evaluate the
distribution function numerically using standard quadrature
techniques.

In order to calculate the density of states function for
$\gamma$-models with a black hole, we can apply the same technique as
for the calculation of the distribution function. If we rewrite
equation~(\ref{dos}) as an integral with the stellar potential as the
integration variable, we obtain
\begin{equation}
    g(\calE)
    =
    16\sqrt{2}\,\pi^2
    \int_{\calE_*}^{\Psi_0}
    \left|
    r^2\frac{\txd r}{\txd\Psi_*}
    \right|
    \sqrt{\omega(\Psi_*)-\omega(\calE_*)}\,
    \txd\Psi_*.
\label{dosBH}
\end{equation}
This integral shows many similarities with the integral in the
expression~(\ref{defQ}), and the prospects to find an analytical
solution are very similar. One can demonstrate that the density of
states function can be written in terms of elliptic integrals for
$\gamma=1$ and $\gamma=\tfrac{3}{2}$, and completely in terms of elementary
functions for $\gamma=\tfrac{5}{2}$. For all other values of $\gamma$,
including the Jaffe model, the density of states function cannot be
expressed analytically.

\subsection{Asymptotic expansions}

\begin{figure*}
\centering
\includegraphics[clip,bb=43 50 351 784,angle=-90,width=\textwidth]{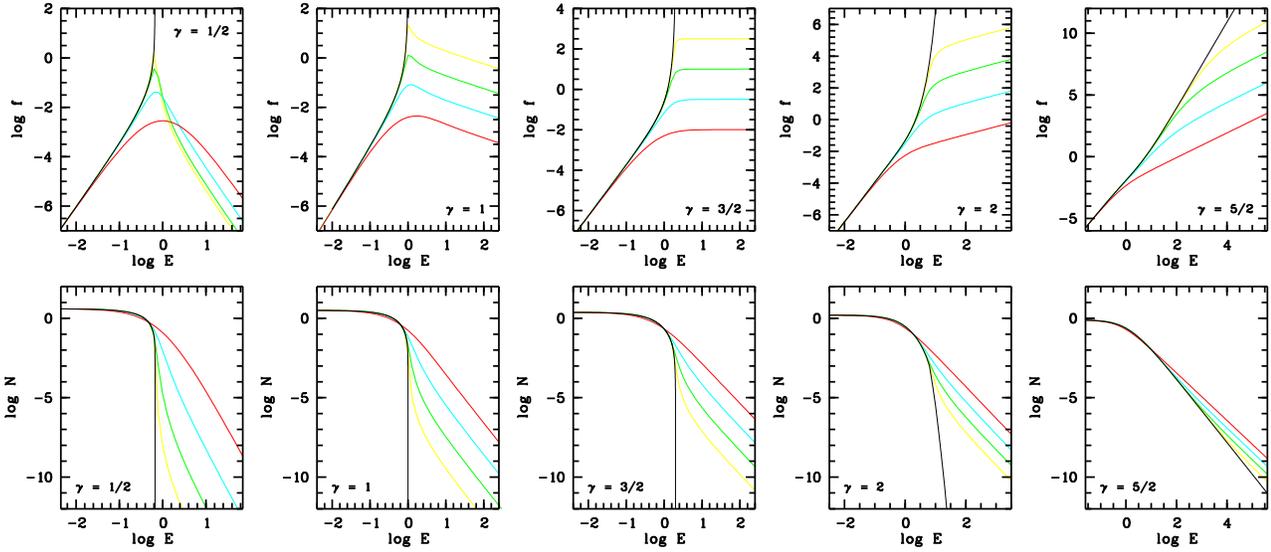}
\caption{The distribution function (top row) and differential energy
distribution (bottom row) for $\gamma$-models without and with a
central black hole. The value of $\gamma$ is indicated in each
panel. The black curves correspond to the self-consistent models, the
colored curves correspond to models with various black hole masses:
$\mu=0.001$ (yellow), $\mu=0.01$ (green), $\mu=0.1$ (cyan) and $\mu=1$
(red).}
\label{dfdos.eps}
\end{figure*}

We can get a better insight into the behaviour of the distribution
function and the DED of the $\gamma$-models with a central black hole
by studying the asymptotic behaviour. In the limit
$\calE\rightarrow0$, i.e.\ in the outer regions of the system, a black
hole has no effect on the dynamics of the $\gamma$-models, as the
asymptotic expansion of the potential $\Psi\rightarrow r^{-1}$ is
independent of $\mu$. The behaviour of the distribution function and
the DED in the low binding-energy limit is
\begin{gather}
    f(\calE)
    \sim
    \frac{4\sqrt{2}(3-\gamma)}{5\pi^3}\,\calE^{5/2},
    \\
    \calN(\calE)
    \sim
    \frac{8(3-\gamma)}{5}.
\label{fsmallE}
\end{gather}
More interesting is the asymptotic behaviour of the distribution
function and the differential energy distribution in the high
binding-energy limit. To find asymptotic expansions for the
self-consistent $\gamma$-models, we must separately consider the
models with a modest and those with a strong central density cusp. The
self-consistent $\gamma$-models with a modest density cusp $\gamma<2$
have a finite potential well $\Psi_0=(2-\gamma)^{-1}$, and after some
algebra one finds that in the limit $\calE\rightarrow\Psi_0$,
\begin{gather}
    f_*(\calE)
    \sim
    \frac{\sqrt{2}}{8\pi^{5/2}}\,
    \frac{\gamma\,(3-\gamma)}{(2-\gamma)^{\frac{4-\gamma}{2-\gamma}}}\,
    \frac{\Gamma\left(\frac{6-\gamma}{4-2\gamma}\right)}
    {\Gamma\left(\frac{4-\gamma}{2-\gamma}\right)}\,
    (\Psi_0-\calE)^{-\frac{6-\gamma}{4-2\gamma}},
    \label{fnoBHlargeEweak}
    \\
    \calN_*(\calE)
    \sim
    \frac{2\gamma\,(3-\gamma)}{(2-\gamma)^{\frac{3-2\gamma}{2-\gamma}}}\,
    \frac{\Gamma\left(\frac{6-\gamma}{4-2\gamma}\right)
    \Gamma\left(\frac{3}{2-\gamma}\right)}
    {\Gamma\left(\frac{4-\gamma}{2-\gamma}\right)
    \Gamma\left(\frac{12-3\gamma}{4-2\gamma}\right)}\,
    (\Psi_0-\calE)^{\frac{1}{2-\gamma}}.
    \label{NnoBHlargeEweak}
\end{gather}
The self-consistent models with a central density cusp slope
$\gamma>2$ have an infinitely deep potential well, and for these
models one finds in the limit $\calE\rightarrow\infty$
\begin{gather}
    f_*(\calE)
    \sim
    \frac{\sqrt{2}}{8\pi^{5/2}}\,
    \gamma\,(3-\gamma)\,(\gamma-2)^{\frac{4-\gamma}{\gamma-2}}\,
    \frac{\Gamma\left(\frac{2}{\gamma-2}\right)}
    {\Gamma\left(\frac{2+\gamma}{2\gamma-4}\right)}\,
    \calE^{\frac{6-\gamma}{2\gamma-4}},
    \label{fnoBHlargeEstrong}
    \\
    \calN_*(\calE)
    \sim
    \frac{2\gamma\,(3-\gamma)}{(\gamma-2)^{\frac{2\gamma-3}{\gamma-2}}}\,
    \frac{\Gamma\left(\frac{2}{\gamma-2}\right)
    \Gamma\left(\frac{8-\gamma}{2\gamma-4}\right)}
    {\Gamma\left(\frac{2+\gamma}{2\gamma-4}\right)
    \Gamma\left(\frac{\gamma+1}{\gamma-2}\right)}\,
    \calE^{-\frac{1}{\gamma-2}}.
    \label{NnoBHlargeEstrong}
\end{gather}
Finally, for the Jaffe model $\gamma=2$, the model that separates
these two classes, the expansion in the limit $\calE\rightarrow\infty$
reads
\begin{gather}
    f_*(\calE)
    \sim
    \frac{1}{4\pi^{5/2}}\,{\text{e}}^{2\calE},
    \\
    \calN_*(\calE)
    \sim
    \frac{2\sqrt{6}}{9}\,{\text{e}}^{-\calE}.
\end{gather}
When the $\gamma$-models contain a black hole, the potential well is
infinitely deep for all values of the cusp slope $\gamma$. The
asymptotic behaviour of the distribution function and the differential
energy distribution changes to
\begin{gather}
    f(\calE)
    \sim
    \frac{3-\gamma}{2(2\pi)^{5/2}}\,
    \frac{\Gamma(\gamma+1)}{\Gamma(\gamma-\tfrac{1}{2})}\,
    \mu^{-\gamma}\,\calE^{\gamma-3/2},
    \label{flargeE}
    \\
    \calN(\calE)
    \sim
    \frac{\sqrt{\pi}\,(3-\gamma)}{8}\,
    \frac{\Gamma(\gamma+1)}{\Gamma(\gamma-\tfrac{1}{2})}\,
    \mu^{3-\gamma}\,\calE^{\gamma-4}.
    \label{NlargeE}
\end{gather}
These expressions are valid for all models with a central black hole
$(0<\mu\leq1)$ and for all values of $\gamma>\tfrac{1}{2}$. For the limiting
case $\gamma=\tfrac{1}{2}$, the leading terms from equations~(\ref{flargeE})
and (\ref{NlargeE}) vanish and the asymptotic expansion for
$\calE\rightarrow\infty$ reads
\begin{gather}
    f(\calE)
    \sim
    \frac{\sqrt{8}}{\pi^3}\,\frac{1}{\sqrt{\mu}}\,\calE^{-5/2},
    \\
    \calN(\calE)
    \sim
    4\mu^{5/2}\,\calE^{-5}.
\end{gather}

\subsection{Discussion}

In the top panels of figure~{\ref{dfdos.eps}} we plot the distribution
function of various $\gamma$-models with various black hole masses. A
black hole drastically changes the behaviour of the distribution
function, particularly for the models with a shallow cusp slope
$(\gamma<2)$ which have a finite stellar potential well.  For these
models, the self-consistent distribution function is a strongly
increasing function of energy, which diverges for $\calE \rightarrow
\Psi_0$ according to formula~(\ref{fnoBHlargeEweak}). When a central
black hole is present in these model, stars of all binding energies
populate the galaxy. The behaviour of the distribution function in
this {\em new territory} depends on the cusp slope, as prescribed by
the asymptotic expression~(\ref{flargeE}). For models with
$\gamma<\tfrac{3}{2}$, the distribution function decreases as a
function of binding energy in the neighborhood of the black hole. The
distribution function of such models thus has two regimes: it
converges to zero both in the low and high binding energy limit and
has a maximal value at binding energies around the depth of the
stellar potential well. The larger the black hole mass, the larger the
value of the binding energy where the distribution function becomes
maximal and the smoother the transition between the two regimes. For
the $\gamma=\tfrac{3}{2}$ model, the distribution function becomes
asymptotically flat in the high binding energy limit. For models with
a steeper cusp slope, the distribution function is a monotonically
increasing function of binding energy, and the differences between a
model without and with black hole becomes less pronounced. In
particular for models with $\gamma\geq2$, the presence of a black hole
does not drastically change the behaviour of the distribution
function. Both without and with a central black hole, the distribution
function is a monotonically increasing function of binding energy,
diverging in the high energy limit.
Equations~(\ref{fnoBHlargeEstrong}) and (\ref{flargeE}) show that the
slope of the distribution function in the high energy limit changes
from $(6-\gamma)/(2\gamma-4)$ to $\gamma-\tfrac{3}{2}$ when a black
hole is present.

In the bottom row of figure~{\ref{dfdos.eps}} we plot the differential
energy distribution ${\mathcal{N}}(\calE)$ for the same models as in
the top row. For all models, ${\mathcal{N}}(\calE)$ converges to a
finite value $8(3-\gamma)/5$ in the low binding energies
limit. Typically, ${\mathcal{N}}(\calE)$ is hardly affected for low
binding energies, where it is a decreasing function of increasing
black hole mass. Only at high binding energies, the effect of a black
hole becomes visible, in particular for models with a shallow density
cusp. The differential energy distribution ${\mathcal{N}}_*(\calE)$ of
the models without black hole suddenly drops to zero when $\calE$
approaches the depth of the potential well, in spite of the divergence
of the distribution function [see
equation~(\ref{NnoBHlargeEweak})]. In the presence of a black hole,
where stars can populate orbits with arbitrarily high binding
energies, $\calN(\calE)$ smoothly decreases as $\calE^{\gamma-4}$ in
the high energy limit. For $\gamma$-models with a steeper cusp slope,
in particular the models with an infinitely deep stellar potential
well, the effect on the differential energy distribution is weaker.

Although the effect of a black hole on the shape of the distribution
function is severe, it thus appears that the effect on the
differential energy distribution is rather modest, even for the models
with a finite stellar potential well. In the energy region that was
off limits for the self-consistent models, $\calN(\calE)$ assumes very
low values. Although the addition of a black hole opens up a huge
range for possible binding energies, the number of stars that actually
populate these orbits is thus fairly small. This can be quantified by
calculating the mean binding energy $\langle\calE\rangle$ of stars in
the models. The mean binding energy is defined as
\begin{equation}
    \langle\calE\rangle
    =
    \int_0^\infty \calN(\calE)\,\calE\,\txd\calE.
    \label{meanE}
\end{equation}
A more straightforward way to calculate the mean binding energy is
through the formula
\begin{align}
    \langle\calE\rangle
    &=
    \iiint
    \txd\bfr
    \iiint
    f(\bfr,\bfv)
    \left[\Psi(r)-\tfrac{1}{2}\bfv^2\right]
    \txd\bfv
    \\
    &=
    4\pi\int_0^\infty
    \rho(r)
    \left[\Psi(r)-\tfrac{3}{2}\sigma^2(r)\right]
    r^2\,\txd r.
\end{align}
Using similar algebra as in section~4, we find
\begin{equation}
    \langle\calE\rangle
    =
    \frac{3}{4(5-2\gamma)}
    \left[1+\mu\left(\frac{4-\gamma}{3(2-\gamma)}\right)\right].
\end{equation}
Numerical integration of the equation~(\ref{meanE}) for a selection of
models gives identical results. For a Hernquist model ($\gamma=1$), the
ratio $\langle\calE\rangle/\langle\calE\rangle_*$ equals $1+\mu$, so
the mean binding energy for a model with $\mu=0.001$ is only 0.1\%
larger than the mean binding energy of the self-consistent
model. 

Another characteristic that can be used to quantify the importance of
a central black hole on the energy distribution is the fraction
$\theta$ of stars on orbits with a binding energy larger than the
potential well of the corresponding self-consistent model,
\begin{equation}
    \theta
    =
    \int_{\Psi_0}^\infty
    {\mathcal{N}}(\calE)\,\txd\calE.
\end{equation}
For a $\gamma=1$ model with black hole masses $\mu=0.001$ and
$\mu=0.01$, a numerical integration yields $\theta=1.63\times10^{-4}$
and $\theta=1.81\times10^{-3}$ respectively.

These numbers clearly demonstrate that the effect of a black hole on
the global energy distribution is actually quite small, although the
distribution function when represented as a function of the binding
energy is affected in a very significant way.

\section{The LOSVD}

The LOSVD (also called line profile or velocity profile)
$\phi_\txp(R,v_\parallel)$ is the distribution of line-of-sight
velocities at a given projected radius $R$. From an observer's point
of view, this is definitely one of the most important kinematical
quantities of a galaxy model, because a LOSVD contains all kinematic
information that can be obtained from a galaxy at a given line of
sight and LOSVDs are in principle directly observable.

\begin{figure*}
\centering
\includegraphics[clip,bb=43 50 502 784,angle=-90,width=\textwidth]{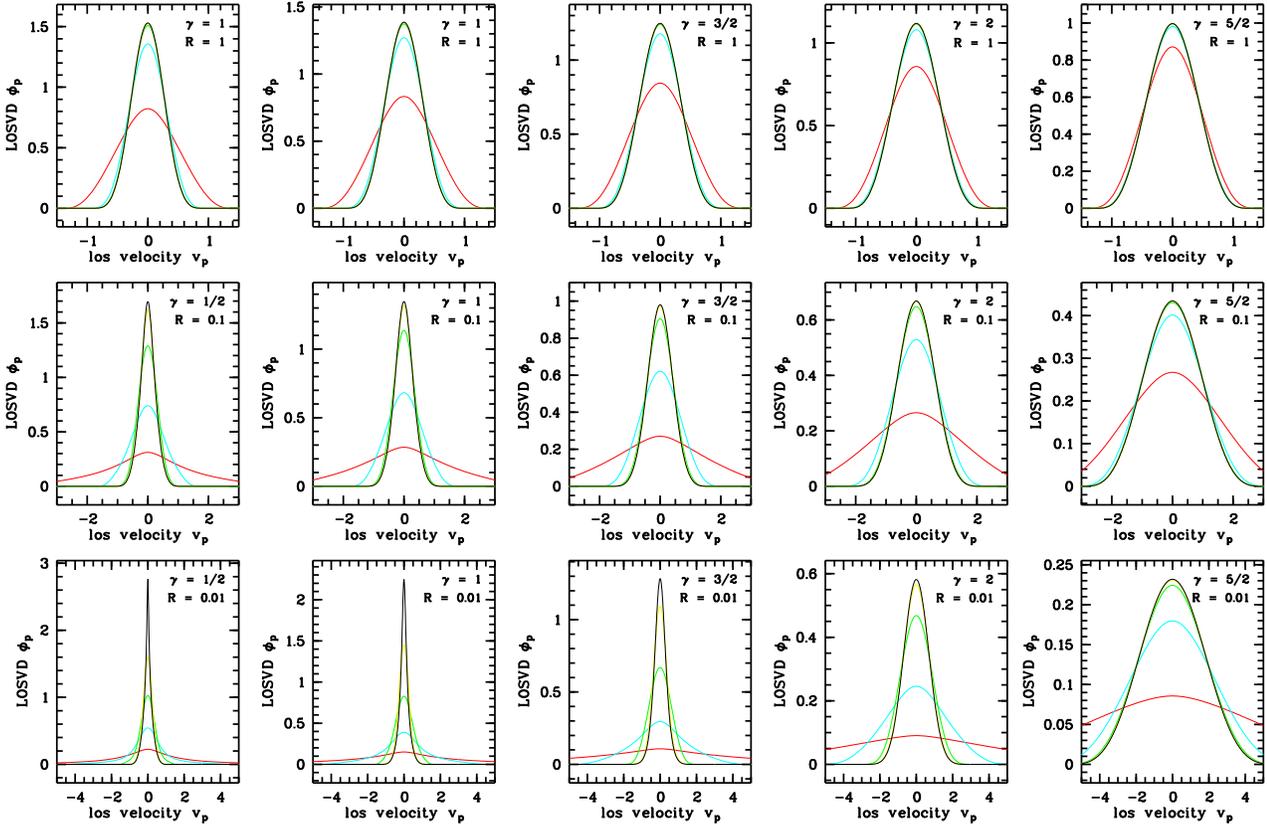}
\caption{The observed LOSVDs for $\gamma$-models without and with a
central black hole. The top panels are the LOSVDs at a projected
radius $R=1$, the middle panels correspond to $R=0.1$ and the bottom
panels are the LOSVDs at $R=0.01$. The value of $\gamma$ is indicated
each panel and the colour code is the same as in
Figure~{\ref{dfdos.eps}}.}
\label{losvd.eps}
\end{figure*}

\subsection{Calculation of the LOSVDs}

The LOSVD can be found through the formula
\begin{equation}
    \phi_\txp(R,v_\parallel)
    =
    \frac{2}{I(R)}
    \int_R^\infty \frac{\rho(r)\,\phi(r,v_\parallel)\,r\,\txd
    r}{\sqrt{r^2-R^2}},
\label{losvd}
\end{equation}
where the function $\phi(r,v_\parallel)$ represents the spatial
LOSVD. In general, the spatial LOSVD $\phi(\bfr,\bfk,v_\parallel)$
describes the distribution of line-of-sight velocities at a position
$\bfr$ in an arbitrary direction $\bfk$. It can be found by
integrating the distribution function over the two velocity components
perpendicular to $\bfk$ and normalizing the resulting distribution
\begin{equation}
    \phi(\bfr,\bfk,v_\parallel)
    =
    \frac{1}{\rho(r)}
    \iint f(\bfr,\bfv)\,
    \txd\bfv_\perp.
\label{slosvd}
\end{equation}
For general anisotropic systems, these integrations are usually very
cumbersome, and an analytical evaluation of the spatial LOSVD can only
be obtained for a limited number of models (e.g.\ Carollo et al.~1995;
Baes \& Dejonghe~2002b). For isotropic models, however, the spatial
LOSVD is independent of the direction $\bfk$ and can be written as
$\phi(r,v_\parallel)$. The expression~(\ref{slosvd}) can then be
transformed to
\begin{equation}
        \phi(r,v_\parallel)
        =
        \frac{2\pi}{\rho(r)}
        \int_0^y f(\calE)\,\txd\calE,
\end{equation}
where $y = \Psi(r)-v_\parallel^2/2$. If we substitute the Eddington
formula~(\ref{eddington}) into this equation, we obtain
\begin{align}
        \phi(r,v_\parallel)
        =
        \frac{1}{\sqrt{2}\pi\,\rho(r)}
        \int_0^y
        \frac{\txd\tilde\rho}{\txd\Psi}\,\frac{\txd\Psi}{\sqrt{y-\Psi}}.
        \label{phirv}
\end{align}
Comparing this expression with the Eddington formula
(\ref{eddington}), we see that the spatial LOSVD of an isotropic
dynamical model can be calculated analytically if the same is true for
the distribution function. In fact, the spatial LOSVD is obtained
almost automatically during the calculation of the distribution
function. This means that the spatial LOSVD of the self-consistent
$\gamma$-models can generally be expressed in terms of hypergeometric
functions, and in terms of elementary functions for $\gamma=2\pm1/n$
with $n$ a natural number. The spatial LOSVD of the black hole
dominated $\gamma$-models can also be expressed in terms of
hypergeometric functions, and in terms of standard functions for
integer and half-integer values of $\gamma$. To calculate the spatial
LOSVD for the $\gamma$-models with a black hole, we use a similar
method as for the calculation of the distribution function. Changing
the integration variable from $\Psi$ to $\Psi_*$ in
equation~(\ref{phirv}) we find
\begin{equation}
        \phi(r,v_\parallel)
        =
        \frac{1}{\sqrt{2}\pi\,\rho(r)}\,
        {\mathcal{Q}}(y_*),
\end{equation}
with $y_*=\omega^{-1}(y)$ and the function ${\mathcal{Q}}$ defined in
equation~(\ref{defQ}). For the models with $\gamma=1$,
$\gamma=\tfrac{3}{2}$ and $\gamma=\tfrac{5}{2}$, both the inverse of
the mapping function $\omega$ and the function ${\mathcal{Q}}$ can be
expressed analytically, such that the spatial LOSVD can be written
completely in analytical form (including elliptic integrals for the
$\gamma=1$ and $\gamma=\tfrac{3}{2}$ models). For the other values of
$\gamma$, both of these operations must be done numerically.

\begin{figure*}
\centering
\includegraphics[clip,bb=43 50 351 784,angle=-90,width=\textwidth]{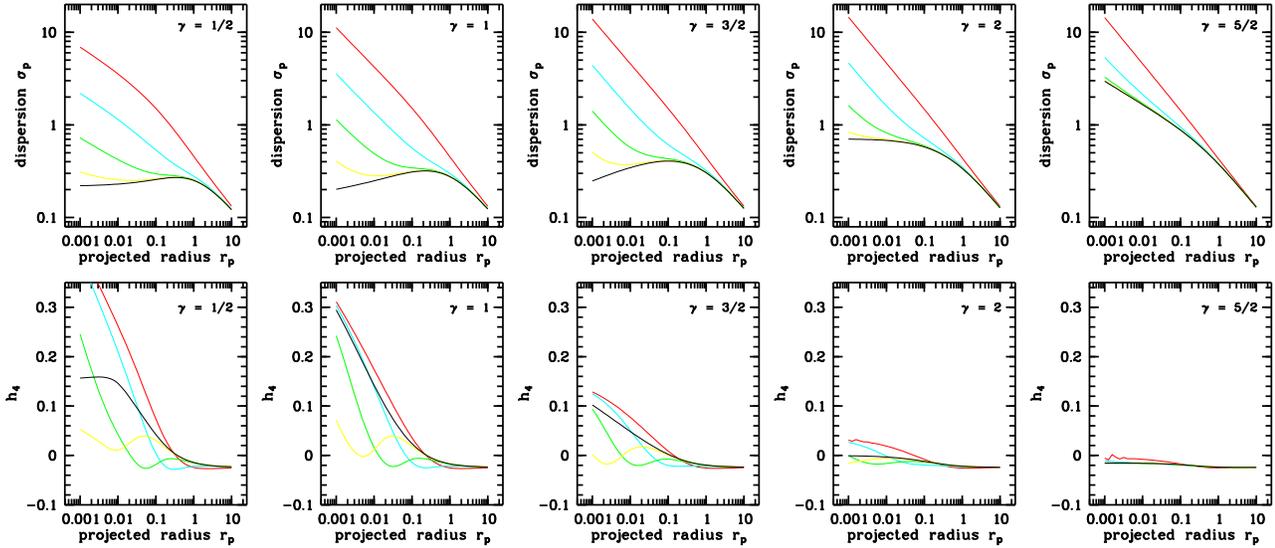}
\caption{The moments of the LOSVDs for $\gamma$-models without and
with a central black hole. The top row shows the projected dispersion
profile $\sigma_\txp(R)$, the bottom row shows the Gauss-Hermite shape
parameter $h_4(R)$. The value of $\gamma$ is indicated in each panel
and the colour code is the same as in Figure~{\ref{dfdos.eps}}.}
\label{mom.eps}
\end{figure*}

Summarizing, the calculation of the LOSVDs requires only one single
quadrature for all self-consistent and black hole dominated models,
and for the $\gamma$-models with a black hole if $\gamma=1$,
$\gamma=\tfrac{3}{2}$ or $\gamma=\tfrac{5}{2}$. For the remaining
$\gamma$-models containing a central black hole, the calculation of
the LOSVDs requires a double quadrature. We calculated the LOSVDs with
the Monte Carlo integration routines built into the SKIRT code (Baes
et al.~2003b).

\subsection{Discussion}

In figure~{\ref{losvd.eps}} we plot a number of LOSVDs for a set of
$\gamma$-models for various values of the black hole mass $\mu$. As
expected, the influence of a (realistic) black hole is negligible at
large projected radii. At gradually smaller projected radii the
influence of a black hole becomes more important. The signature of the
presence of the black hole is primarily a broadening of the LOSVD. The
degree of the broadening depends on the black hole fraction and on the
central slope.

A quantitative investigation of the effects of a black hole on the
LOSVDs is more easy when we study the moments of the LOSVDs. The first
and second-order moments of the LOSVDs are nothing but the mean
rotation and the projected dispersion. Rather than the true
higher-order moments, one usually utilizes the coefficients $h_i$ from
a Gauss-Hermite expansion to further characterize the shape of the
LOSVDs (Gerhard~1993; van der Marel \& Franx~1993). In
figure~{\ref{mom.eps}} we plot the projected velocity dispersion
profile and the $h_4$ profile for a set of $\gamma$-models without and
with black holes of various masses (all odd moments are zero).

The projected dispersion profile of models with a shallow central
density slope are most sensitive to a black hole. As a result of the
presence of a black hole, the stars in the very centre of the galaxy
will obtain large velocities, including velocities which are not
possible in the finite central potential wells of the self-consistent
models. This causes a dramatic increase in the velocity dispersion at
small radii, resulting in the usual $R^{-1/2}$ divergence. Note that
this divergence is maintained in spite of the smoothing effect of the
projection along the line of sight: the projected dispersion is a
weighted integral of the intrinsic dispersion along the line of sight,
such that the projected dispersion at small projected radii contains a
significant contribution from stars at large radii.

The projected dispersion profile for models with a steep cusp slope is
much less affected by a central black hole. Even without a black hole,
these models already have steep stellar potential wells where stars
can obtain arbitrarily high velocities. The limit case is the
(degenerate) model with $\gamma\rightarrow3$, where the
self-consistent potential is already a point potential, and the
addition of a black hole does not alter the kinematical structure of
the galaxy at all.

The effect of a black hole on the $h_4$ shape parameter of the LOSVD
is less straightforward to interpret. For the models with a shallow
cusp, the central black hole does significantly affect the $h_4$
profile, but there is no clear trend with increasing black hole
mass. For the models with a steep density cusp, the effect of a black
hole on the $h_4$ profile is negligible. This trend, or rather the
non-existence of a clear trend, contrasts with the signature of a dark
matter halo on the observed kinematics at large radii. In elliptical
galaxies, which lack a ubiquitous and straightforward tracer
population such as the H{\sc i} gas in spiral galaxies, the stellar
kinematics are one of the most important tracers for dark matter. The
most obvious signature of dark matter on the stellar kinematics is the
behaviour of the projected velocity dispersion at large radii: dark
matter causes the $\sigma_\txp$ profile to remain flat or to decrease
more slowly than expected from the photometry alone (Saglia, Bertin \&
Stiavelli~1992; Saglia et al.~1993). This observational signature
suffers from a mass-anisotropy degeneracy, however: a tangential
stellar orbital structure can cause a similar signature
(Gerhard~1993). The key to discriminate between these possibilities is
the behaviour of the $h_4$ profile. Carollo et al.~(1995) studied in
some detail the effect of a dark matter halo on the $h_4$ shape
parameter. They found that, both for isotropic and radially
anisotropic models, the $h_4$ parameter increases at large projected
radii due to the presence of a dark matter halo. On the other hand, a
tangential orbital structure causes a negative $h_4$ profile at large
radii.  The measurement of the projected dispersion and the $h_4$
profile at large radii would therefore in principle suffice to
constrain the dark matter content in ellipticals (Rix et al.~1997;
Gerhard et al.~1998; Kronawitter et al.~2000), although systematic
effects caused by dust, non-sphericity etc.\ might further complicate
this issue (Baes \& Dejonghe~2001; Sanchis, {\L}okas \&
Mamon~2004). There is apparently no analogy for this situation for the
detection of black holes in galactic nuclei: as the signature of a
central black hole on the $h_4$ profile does not follow a clear trend,
the projected velocity dispersion is by far the most obvious
observational signature by which a black hole can be detected in
non-rotating stellar nuclei.

\section{Discussion and conclusion}

Galactic nuclei observed in real galaxies show a variety of structure,
with central density cusps ranging from flat cores to steep cusps. The
main goal of this paper was to construct simple dynamical models for
spherical systems with a central black hole, reflecting the variety of
central structure observed in real galactic nuclei. We have performed
a detailed study of a two-parameter family of dynamical models based
on the family of $\gamma$-models introduced by Dehnen~(1993) and
Tremaine et al.~(1994). The family contains as parameters the slope of
the central density cusp $\gamma$ and the ratio $\mu$ of the central
black hole mass to the total mass of the system. By varying these
parameters, we have been able to study a very wide range of models,
going from weakly cusped models to very centrally concentrated models
with an infinite stellar potential well, and ranging from
self-consistent models without black hole to systems where the
dynamical structure is completely dominated by the central black
hole. We have only considered models with an isotropic dynamical
structure, which do not cover the range of orbital structure observed
in real galactic nuclei. It is possible to generalize the isotropic
models presented in this paper to anisotropic models with a constant
anisotropy or with an Osipkov-Merritt type distribution function
(Osipkov 1979; Merritt~1985). For such models, the calculation of the
most important dynamical properties is not much more demanding than
for isotropic models. The construction of general anisotropic
dynamical models is much more complicated, however, and falls beyond
the scope of this paper.

For this two-parameter family of isotropic dynamical models, we have
calculated the most important kinematical quantities, such as the
intrinsic and projected velocity dispersions, the total energy budget,
the distribution function, the differential energy distribution and
the LOSVDs. Many of these quantities could be calculated completely
analytically for all values of $\gamma$ and $\mu$. For the models
without black hole and the black hole dominated models, the
distribution function, the differential energy distribution and the
spatial LOSVDs can be calculated completely analytically. The same is
true for three distinct $\gamma$-models with a central black hole --
these three models each form an analytical one-parameter family with
the central black hole mass as an explicit parameter. Although these
three models differ only by the slope of the central density profile,
their kinematical properties are very different. We therefore believe
that the set of $\gamma$-models discussed in this paper, and in
particular the three completely analytical models, answer the need for
dynamical models which are on the one hand relatively simple and on
the other hand reflect the range of central structure observed in the
nuclei of real galaxies.

As most of the kinematics can be calculated completely analytically,
the presented models allow to investigate the effect of a central
black hole on the kinematics of galactic nuclei. In general, the
effect of a black hole depends on the central density cusp (the
parameter $\gamma$). Models with a steep density cusp are least
affected by the presence of a black hole. They already have a steep
and infinitely deep potential well and a strong central density
concentration, and the kinematical effect of a black hole is quite
marginal. The distribution function remains a monotonically increasing
function of binding energy; only the slope in the high-energy limit is
altered by the presence of a black hole. Models with a shallow central
density cusp on the other hand are more strongly affected by the
presence of a black hole. They have a less concentrated density
profile and therefore only a finite central potential well. The
influence of a black hole on the distribution function can be
important. For example, models with $\gamma<\tfrac{3}{2}$ have distribution
functions which tend to zero both in the low and high binding energies
regime. The differential energy distribution, which gives a better
physical insight in the orbital structure of the system, is less
affected by the presence of a black hole. For realistic black hole
masses, the shift in the mean binding energy of the stars or the
fraction of stars on orbits with binding energies exceeding the depth
of the stellar potential well is marginal. Nevertheless, the different
behaviour of the distribution function is important, as it bears
direct consequences for the stability of the $\gamma$-models with a
black hole. A monotonically rising distribution function as a function
of binding energy is a sufficient condition for stability against
radial and non-radial perturbations. This means that the
$\gamma$-models with a black hole with $\gamma\geq\tfrac{3}{2}$ are stable. The
models with a shallower cusp slope have a decreasing distribution
function in the high binding energy limit, and their stability is
hence uncertain. This interesting issue can only be investigated in a
detailed N-body or linear mode analysis studies.

Finally, we can wonder whether these three models are unique or
whether other simple dynamical models could easily be found for which
most of the kinematical properties can be expressed analytically. It
is rather straightforward to constructing models with a black hole in
which the (intrinsic and/or projected) velocity dispersions can be
expressed analytically, as the velocity dispersion is just a linear
function of the black hole mass. The construction of models where more
complicated dynamical properties such as the distribution function and
the spatial LOSVDs can be expressed analytically is more difficult. A
direct integration of the Eddington equation does not seem the most
obvious way to proceed [however, see Baes \& Dejonghe (2004) for a
case where this is doable]. A more promising path is the idea of
mapping the total potential onto the stellar potential, proposed by
Ciotti~(1996) and adapted by us in the present paper. Nevertheless,
the conditions on the mapping function in order to allow an analytical
evaluation of the distribution function are quite stringent, and they
probably do not apply for a large set of models. Out of the range of
$\gamma$-models, we found that only three different models satisfy
these conditions. We also searched for other models in the large
family of the so-called $(\alpha,\beta,\gamma)$-models (Zhao~1996), an
extension of the $\gamma$-models, but no other models satisfied the
necessary conditions.

\begin{acknowledgement}
The main part of this work was done when MB was a Postdoctoral Fellow
of the Fund for Scientific Research Flanders (FWO). The numerical
calculations were performed using the SKIRT cluster at the
Universiteit Gent, for which MB obtained an FWO research grant. The
FWO is kindly acknowledged for the financial support. The referee, Tim
de Zeeuw, is acknowledged for helpful suggestions.
\end{acknowledgement}

\appendix

\section{The function $W_\gamma(r)$}

The function $W_\gamma(r)$ is defined in equation~(\ref{defW}) as
\begin{equation}
    W_\gamma(r)
    =
    \frac{3-\gamma}{4\pi}
    \int_r^\infty
    \frac{r^{1-2\gamma}\txd r}{(1+r)^{7-2\gamma}},
\label{defW2}
\end{equation}
where it is assumed that $r$ is a real number and the parameter
$\gamma$ assumes values between 0 and 3.  Formally, it can
conveniently be expressed in terms of the incomplete Beta function or
the hypergeometric function. It can also be written completely in
terms of elementary functions. If $\gamma$ is not equal to $1$,
$\tfrac{3}{2}$, $2$ or $\tfrac{5}{2}$, we obtain an algebraic
expression
\begin{equation}
    W_\gamma(r)
    =
    \frac{3-\gamma}{4\pi}
    \sum_{j=0}^4
    \binom{4}{j}\,
    \frac{(-1)^j}{2+j-2\gamma}\,
    \left[1-\left(\frac{r}{1+r}\right)^{2+j-2\gamma}\right].
\end{equation}
For the four special values of $\gamma$, we can obtain a closed
expression for $W_\gamma(r)$ through direct integration of
equation~(\ref{defW2}) or through application of the reduction
formulae of hypergeometric functions. The results are
\begin{gather}
    W_1(r)
    = \frac{1}{2\pi}\ln\left(\frac{1+r}{r}\right)
    - \frac{25+52\,r+42\,r^2+12\,r^3}{24\pi\,(1+r)^4},
    \\
    W_{3/2}(r)
    = -\frac{3}{2\pi}\ln\left(\frac{1+r}{r}\right)
    + \frac{3+22\,r+30\,r^2+12\,r^3}{8\pi\,r\,(1+r)^3},
    \\
    W_2(r)
    = \frac{3}{2\pi}\ln\left(\frac{1+r}{r}\right)
    + \frac{1-4\,r-18\,r^2-12\,r^3}{8\pi\,r^2\,(1+r)^2},
    \\
    W_{5/2}(r)
    = -\frac{1}{2\pi}\ln\left(\frac{1+r}{r}\right)
    + \frac{1-2\,r+6\,r^2+12\,r^3}{24\pi\,r^3\,(1+r)}.
\end{gather}

\section{The function $Q(x)$ for some selected models }

\subsection{The $\gamma=1$ model}

For the Hernquist model we find a simple augmented density,
\begin{equation}
    \tilde\rho(\Psi_*)
    =
    \frac{1}{2\pi}\,\frac{\Psi_*^4}{1-\Psi_*},
    \label{adH}
\end{equation}
and for the mapping function we find
\begin{equation}
    \omega(\Psi_*)
    =
    \Psi_*\left(1+\frac{\mu\Psi_*}{1-\Psi_*}\right),
\label{omegaH}
\end{equation}
If we substitute the expressions~(\ref{adH}) and~(\ref{omegaH}) into
equation~(\ref{defQ}), we obtain
\begin{equation}
    {\cal{Q}}(x)
    =
    \frac{1}{2\pi\sqrt{1-\mu}}
    \int_0^x
    \frac{\Psi_*^3\,(3\Psi_*-4)\,\txd\Psi_*}
    {\sqrt{(x-\Psi_*)\,(1-\Psi_*)^3\,(\zeta-\Psi_*)}},
    \label{QH}
\end{equation}
where $\zeta = 1+\mu/(1-\mu)(1-x)$. The integrand in this expression
is basically the combination of a rational function and the square
root of a cubic polynomial. The function ${\mathcal{Q}}$ (and hence
the distribution function) can therefore be expressed completely in
terms of elliptic integrals. An explicit form requires a reduction of
this elliptic integral to a combination of the standard elliptic
integrals. For the Hernquist model, one finds that the distribution
can be written as
\begin{equation}
  f(\calE)
  =
  A_1+A_2\mathbb{F}(\phi,k)+A_3\mathbb{E}(\phi,k),
\end{equation}
where $A_i$ are algebraic functions of $\calE$ and $\mu$, $\mathbb{F}$
and $\mathbb{E}$ are Legendre's incomplete elliptic integrals of the
first and second kind, and the arguments of the elliptic integrals are
\begin{gather}
  \phi = \arcsin\sqrt{\calE},
  \\
  k = \left[1+\frac{(1-\mu)(1-\calE)^2}{\mu}\right]^{-1/2}.
\end{gather}
For the calculation of the density of states, similar formulae
apply. Practically, it is easiest to use a symbolic mathematical
package such as Maple or Mathematica. These packages return an
explicit form for most elliptic integrals, and the expressions can be
converted immediately to Fortran or C code using internal conversion
routines.

\subsection{The $\gamma=\tfrac{3}{2}$ model}

For $\gamma=\tfrac{3}{2}$ we obtain
\begin{gather}
    \tilde\rho(\Psi_*)
    =
    \frac{3}{256\pi}\,\frac{(4-\Psi_*)^4\Psi_*^4}{(2-\Psi_*)^3},
    \label{adD}
    \\
    \omega(\Psi_*)
    =
    \Psi_*\left[1+\frac{\mu\,\Psi_*\,(3-\Psi_*)}{(2-\Psi_*)^2}\right].
    \label{omegaD}
\end{gather}
Substitution of (\ref{adD}) and (\ref{omegaD}) into
formula~(\ref{defQ}) yields
\begin{multline}
    {\mathcal{Q}}(x)
    =
    \frac{3}{256\pi\sqrt{1-\mu}}
    \times
    \\
    \int_0^x
    \frac{\Psi_*^3\,(4-\Psi_*)^3\,
    (32-20\Psi_*+5\Psi_*^2)\,\txd \Psi_*}
    {(2-\Psi_*)^3\,
    \sqrt{(x-\Psi_*)\,(\zeta_--\Psi_*)\,(\zeta_+-\Psi_*)}},
\label{calQDehnen}
\end{multline}
where the roots of the cubic polynomial under the square root are
\begin{equation}
    \zeta_\pm
    =
    2 + \frac{2\mu}{(1-\mu)\,(2-x)^2}
    \left[1\pm\sqrt{1-\frac{(1-\mu)\,(2-x)^3}{\mu}}\right].
\label{xipm}
\end{equation}
Since the integrand of ${\mathcal{Q}}$ is the product of a rational
function and the square root of a cubic polynomial, the integral can
be written in terms of elliptic integrals.

\subsection{The $\gamma=\tfrac{5}{2}$ model}

The $\gamma=\tfrac{5}{2}$ model has an augmented density very similar
to that of the $\gamma=\tfrac{3}{2}$ model,
\begin{equation}
    \tilde\rho(\Psi_*)
    =
    \frac{1}{256\pi}\,\frac{(4+\Psi_*)^4\Psi_*^4}{(2+\Psi_*)^3}.
    \label{adS}
\end{equation}
The mapping function $\omega(\Psi_*)$ is very simple
\begin{equation}
    \omega(\Psi_*)
    =
    \Psi_*\left(1+\frac{\mu\Psi_*}{4}\right),
\end{equation}
and substitution of these expressions into equation~(\ref{defQ}) gives
\begin{equation}
    {\mathcal{Q}}(x)
    =
    \frac{1}{128\pi\sqrt{\mu}}
    \int_0^x
    \frac{\Psi_*^3\,(4+\Psi_*)^3\,
    (32+20\Psi_*+5\Psi_*^2)\,\txd\Psi_*}
    {(2+\Psi_*)^4\,\sqrt{(x-\Psi_*)(\Psi_*-\zeta)}},
\end{equation}
where $\zeta=-(x+4/\mu)$. As the integrand in this expression is the
product of a rational function and the square root of a quadratic
polynomial, the function ${\mathcal{Q}}$ (and the distribution
function) can be expressed completely in terms of elementary
functions.  Explicit expressions for the distribution function, as
well as for the density of states function, in term of elementary
function can be found in Baes \& Dejonghe (2004).


\begin{thebibliography}{}
\bibitem{BD01} Baes M., Dejonghe H., 2001, ApJ, 563, L19
\bibitem{BD02a} Baes M., Dejonghe H., 2002a, A\&A, 393, 485
\bibitem{BD02b} Baes M., Dejonghe H., 2002b, MNRAS, 335, 441
\bibitem{BD04} Baes M., Dejonghe H., 2004, MNRAS, 351, 18
\bibitem{BBHD03} Baes M., Buyle P., Hau G.\,K.\,T., Dejonghe H., 2003a, MNRAS, 341, L44
\bibitem{BDDSRELSd03} Baes M., et al., 2003b, MNRAS, 343, 1081 
\bibitem{BT87} Binney J., Tremaine S., 1987, Galaxy Dynamics (Princeton University Press, Princeton)
\bibitem{CdZvdM95} Carollo C.\,M., de Zeeuw P.\,T., van der Marel R.\,P., 1995, MNRAS, 276, 1131 
\bibitem{CSM98} Carollo C.\,M., Stiavelli M., Mack J., 1998, AJ, 116, 68
\bibitem{C96} Ciotti L., 1996, ApJ, 471, 68
\bibitem{dBvdMdZ96} de Bruijne J.\,H.\,J., van der Marel R.\,P., de Zeeuw P.\,T., 1996, MNRAS, 282, 909 
\bibitem{D86} Dejonghe H., 1986, Phys. Rep., 133, 217
\bibitem{D87} Dejonghe H., 1987, MNRAS, 224, 13
\bibitem{D89} Dejonghe H., 1989, in Dynamics of dense stellar systems, Cambridge University Press, p.~69
\bibitem{D93} Dehnen W., 1993, MNRAS, 265, 250
\bibitem{DG94} Dehnen W., Gerhard O.\,E., 1994, 268, 1019
\bibitem{E94} Evans N.\,W., 1994, MNRAS, 267, 333
\bibitem{EHdZ97} Evans N.\,W., Hafner R.\,M., de Zeeuw P.\,T., 1997, MNRAS, 286, 315
\bibitem{F02} Ferrarese L., 2002, ApJ, 578, 90
\bibitem{FM00} Ferrarese L., Merritt D., 2000, ApJ, 539, L9
\bibitem{G+96} Gebhardt K. et al., 1996, AJ, 112, 105
\bibitem{G+00} Gebhardt K. et al., 2000, ApJ, 539, L13
\bibitem{Genz+03} Genzel R. et al., 2003, ApJ, 594, 812
\bibitem{G93} Gerhard O.\,E., 1993, MNRAS, 265, 213
\bibitem{GJSB98} Gerhard O., Jeske G., Saglia R.\,P., Bender R., 1998, MNRAS, 295, 197
\bibitem{GTC01} Graham A. W., Trujillo I., Caon N., 2001, AJ, 122, 1707
\bibitem{H59} H\'enon M., 1959, Ann. d'Astrophys., 22, 126
\bibitem{H61} H\'enon M., 1960, Ann. d'Astrophys., 23, 474
\bibitem{H90} Hernquist L., 1990, ApJ, 356, 359
\bibitem{H94} Hiotelis N., 1994, A\&A, 283, 783
\bibitem{J83} Jaffe W., 1983, MNRAS, 202, 995
\bibitem{JdZ01} Jalali M.\,A., de Zeeuw P.\,T., 2001, 328, 511
\bibitem{KR95} Kormendy J., Richstone D., 1995, ARA\& A, 33, 581
\bibitem{KSGB00} Kronawitter A., Saglia R.\,P., Gerhard O., Bender R., 2000, A\&AS, 144, 53
\bibitem{LABDF95} Lauer T. R., Ajhar E. A., Byun Y., Dressler A., Faber S. M., 1995, AJ, 110, 2622
\bibitem{MGT98} Malkan M.\,A., Gorjian V., Tam R., 1998, ApJS, 117, 25
\bibitem{MH03} Marconi A., Hunt L.\,K., 2003, ApJ, 589, L21
\bibitem{MRMP03} Martini P., Regan M.\,W., Mulchaey J.\,S., Pogge R.\,W., 2003, ApJ, 589, 774 
\bibitem{MD01} McLure R.\,J., Dunlop J.\,S., 2002, MNRAS, 331, 795
\bibitem{M85} Merritt D., 1985, AJ, 90, 1027
\bibitem{O79} Osipkov L. P., 1979, Sov. Astron. Lett., 5, 42
\bibitem{PIMF96} Phillips A.\,C., Illingworth G.\,D., MacKenty J.\,W., Franx M., 1996, AJ, 111, 1566
\bibitem{P11} Plummer H. C., 1911, MNRAS, 71, 460
\bibitem{QdZvdMH95} Qian E.\,E., de Zeeuw P.\,T., van der Marel R.\,P., Hunter C., 1995, MNRAS, 274, 602
\bibitem{RHPFS01} Ravindranath S., Ho L.\,C., Peng C.\,Y., Filippenko A.\,V., Sargent W.\,L.\,W., 2001, AJ, 122, 653
\bibitem{RdCvC97} Rix H.-W., de Zeeuw P.\,T., Cretton N., van der Marel R.\,P., Carollo C.\,M., 1997, ApJ, 488, 702
\bibitem{SBS92} Saglia R.\,P., Bertin G., Stiavelli M., 1992, ApJ, 384, 433 
\bibitem{Sea93} Saglia R.\,P.\ et al., 1993, ApJ, 403, 567
\bibitem{SLM04} Sanchis T., {\L}okas E.\,L., Mamon G.\,A., 2004, 347, 1198
\bibitem{Sea04} Scarlata C.\ et al., 2004, AJ, in press (astro-ph/0408435)
\bibitem{SCSdZD02} Seigar M., Carollo C.\,M., Stiavelli M., de Zeeuw P.\,T., Dejonghe H., 2002, AJ, 123, 184
\bibitem{T82} Toomre A., 1982, ApJ, 259, 535
\bibitem{TTFDJvR01} Tran H.\,D., Tsvetanov Z., Ford H.\,C., Davies J., Jaffe W., van den Bosch F.\,C., Rest A.,
2001, AJ, 121, 2928
\bibitem{T+94} Tremaine S. et al., 1994, AJ, 107, 634
\bibitem{vF93} van der Marel R., Franx M., 1993, ApJ, 407, 525 
\bibitem{Z96} Zhao H., 1996, MNRAS, 278, 488
\bibitem{Z00} Zhenglu J., 2000, MNRAS, 319, 1067
\bibitem{ZM02} Zhenglu J., Moss D., 2002, MNRAS, 331, 117
\end{thebibliography}
\end{document}